\documentstyle[11pt,newpasp,twoside,epsf]{article}
\def\edcomment#1{\iffalse\marginpar{\raggedright\sl#1\/}\else\relax\fi}
\reversemarginpar

\begin{document}
\title{All Sky Doppler Extrasolar Planet Surveys with a Multi-object
Dispersed Fixed-delay Interferometer}
\author{Jian Ge, Suvrath Mahadevan, Julian van Eyken, Curtis DeWitt}
\affil{Department of Astronomy \& Astrophysics, Penn State University University Park, PA 16802}
\author{Stuart Shaklan}
\affil{Jet Propulsion Laboratory, M/S 301-486, 4800 Oak Grove Dr. Pasadena, CA
91109}

\begin{abstract} Characterization of extrasolar planetary systems
requires a radial velocity (RV) survey for planets around hundreds of thousands of
nearby stars of all spectral types over next ten years.
This survey will be  extremely difficult to conduct using 
current high resolution echelle spectrometer due to its single object observing
mode and low instrument throughput. 
Here we propose to use a high throughput multi-object dispersed
fixed-delay interferometer  for the survey. This instrument,
a combination of  a fixed-delay interferometer
with a moderate resolution spectrometer,
is completely different from current 
echelle spectrometers. Doppler 
RV  is measured through monitoring interference fringe  shifts 
of stellar absorption 
lines over a broad band. Coupling this  multi-object
instrument with a wide field telescope (a few
degree, such as Sloan and WIYN) and UV, visible and near-IR
detectors will allow to simultaneously obtain hundreds of
stellar fringing spectra for searching for planets. The RV survey speed can be 
 increased by more than 2 orders of magnitude over that for the echelles.

A prototype dispersed
fixed-delay interferometer has been observed at the Hobby-Eberly
9m and Palomar 5m  telescopes in 2001 and demonstrated
photo noise limited Doppler precision with Aldebaran. Our
recent observations at the KPNO 2.1m telescope in 2002 demonstrate
a short term Doppler precision of $\sim$ 3 m/s with  $\eta$ Cas (V = 3.5), a
RV stable star and also obtained a RV curve for 51 Peg. (V = 5.5), 
confirming previous planet detection with an independent RV
technique. The total measured detection efficiency including the sky,
telescope and fiber transmission losses, the instrument and iodine
transmission losses and detector quantum efficiency is 3.4{\%}
under 1.5 arcsec seeing conditions, which is comparable to all of
the echelle spectrometers for planet detection.

\end{abstract}

\section{Introduction} Since the discovery of 51 Peg B (Mayor {\&} Queloz 1995), 
the number of known planets has gone up dramatically and today more than 100
such companions are known. Most of these systems have been
discovered using high resolution echelle spectrographs. These
instruments are routinely achieving a radial velocity (RV) precision of
$\sigma $ = 5-15 m/s and even as low as 3 m/s in the best cases
(Butler et al. 1996; Baranne et al. 1996; Vogt et al. 2000).
 The echelle spectrographs
themselves are large and expensive, with a complicated point
spread function which has to be modeled appropriately. The high
resolution along with the large wavelength coverage is paid for by
the huge costs, the complexity of the instrument and the low
throughput of around 1-4{\%} (Vogt et al. 2000; D'Odorico et al.
2002). The observational challenge now lies in detecting as many
planetary systems as possible to understand the statistical
distribution of planetary masses and distances from the host star.
Detecting a large sample of planets will also help in
understanding the physical processes underlying planetary
formation. A larger sample of stars needs to be surveyed with
velocity precision less than 10 m/s to make this possible. Such a
task will need large amounts of time on telescopes with current
echelles and this is a significant challenge. The next
generation of RV survey instruments must have higher
efficiencies, should be able to do multi-object RV
surveys and should extend the wavelength range to near IR and near UV.
The development of our instrument, a multiple object dispersed
fixed-delay interferometer, a new generation RV
instrument, has been driven in part by these needs.

\section {Principle and Unique Properties of Fixed-delay Interferometer}
The use of a fixed-delay interferometer for Doppler RV measurements
is completely different from the current echelle approach. Instead
of measuring the absorption line centroid shifts in the echelle
approach, the RV is measured through monitoring interference
fringe shifts (Ge 2002). The original idea for using a
fixed-delay interferometer for high precision Doppler RV
measurements was proposed by  two groups  (Gorskii \& Lebedev
1977; Beckers \& Brown 1978). This interferometer with a narrow
bandpass has been successfully used for very high Doppler
precision measurements of the sun ($\sim $ 3 m/s, Kozhevatov et
al. 1995, 1996; sub m/s precision for the GONG measurements,
Harvey 2002 private communication). The concept of combining of a
fixed-delay interferometer with a moderate resolution spectrometer, or a 
post-disperser, for broad band operations for high precision stellar Doppler
measurements was proposed by Dave Erskine at LLNL in 1997. The
initial lab experiments and telescope observing with a prototype
demonstrated its feasibility (Erskine {\&} Ge 2000; Ge et al.
2002a). A theory for this new instrument concept was developed by
Jian Ge (Ge 2002). In this interferometer approach, the instrument
response, determined by the two beam interference, is a simple and
well-defined sinusoidal function. For comparison, the echelle
response is considerably more complex due to the interference
among thousands of divided beams from grating grooves.

In a fixed-delay interferometer (FDI), a fixed optical delay, $d$, is applied
to one of the beams. Therefore, the interference happens at very high
interference order, $m$, determined by $m = \frac{d}{\lambda }$, where
\textit{$\lambda $} is the operating wavelength. The Doppler RV motion will 
shift the fringes
of stellar absorption lines to neighboring orders. The corresponding Doppler
velocity shift is

\begin{equation}
\label{eq1}
\Delta \upsilon = \frac{c\lambda }{d}\Delta m = \frac{c\lambda
}{d}\frac{\Delta \phi }{2\pi },
\end{equation}

\noindent
where \textit{$\Delta \phi $} is the measured phase shift of a fringe. If the absorption line
density is constant over the observed band, then the total observed Doppler
error is

\begin{equation}
\label{eq2}
\sigma _{t,fringe,ob} \approx \frac{1}{\sqrt {N_o } }\frac{1.1c\lambda }{D_o
l_c \sqrt {F_o } } \approx \sigma _{t,fringe,I}
\end{equation}

\noindent
where $N_{o}$ is the total number of absorption lines covered by the array,
$D_{o}$ is the observed absorption line depth, $l_{c }$\textit{= $\lambda $}$^{2}$\textit{/$\Delta \lambda $}$_{I}$ is the
coherence length of the interferometer beam for a bandwidth of \textit{$\Delta \lambda $}$_{I}$, the
intrinsic width of typical stellar absorption lines, $F_{o}$ is the observed
flux within each observed fringe (or absorption line), and $\sigma
_{t,fringe,I} $ is the intrinsic Doppler precision at infinite spectral
resolution of a post-disperser (see Ge 2002 for details). This indicates
that {\it the Doppler sensitivity of the FDI is independent of the
spectral resolving power of the post-disperser}, contrary to echelle
spectroscopy. In the echelle, the total observed Doppler
error is 

\begin{equation}
\label{eq4}
\sigma _{t,echelle,ob} \approx \frac{1}{\sqrt {N_o } }\frac{1.1c\Delta
\lambda _o }{D_o \lambda \sqrt {F_o } } \approx \left( {\frac{\Delta \lambda
_o }{\Delta \lambda _I }} \right)\sigma _{t,echelle,I} \approx \left(
{\frac{\Delta \lambda _o }{\Delta \lambda _I }} \right)\sigma _{t,fringe,ob},
\end{equation}

\noindent where the observed line FWHM is $\Delta \lambda _o =
\sqrt {\Delta \lambda _I^2 + \Delta \lambda _e^2 } ,$ and
\textit{$\Delta \lambda $}$_{e}$ is the FWHM of the echelle
response, $\sigma_{t,echelle,I} $ is the intrinsic Doppler precision 
at infinite spectral
resolution of an echelle. {\it The Doppler error in the echelle approach
strongly depends on the echelle resolving power and stellar
intrinsic line width} (Bouchy et al. 2001; Ge 2002).
 For a solar
type star with absorption lines of FWHM $\sim $ 5 km/s (Dravins
1987), at moderate resolution (such as \textit{$\Delta \lambda
$}$_{o}$\textit{$\sim $ 10$\Delta \lambda $}$_{I}$, or $R 
\sim $ 6000), the echelle approach has $\sim $ 10 times higher
Doppler error than the interferometer approach. At high
resolution, such as $R \sim $ 60,000, being used for
planet detection (e.g., Vogt et al. 1994; D'Odorico et al. 2000),
the Doppler sensitivity for a solar type star is still $\sim $ 1.4
times worse than the interferometer.

The independence of Doppler sensitivity from the post-disperser resolving
power in the interferometer approach opens up new possibilities for RV
studies. The use of low resolution but high efficiency post-dispersers can
significantly boost the overall detection efficiency, dramatically reduce
the instrument size and cost and allow single dispersion order operations
for multiple object observations. Full sky coverage for an RV survey for
planets becomes possible with wide field telescopes. {\it Multiple object
capability is one of the most significant advantages for this interferometer
approach.} The simple and stable response function in the interferometer
approach leads to potential low systematic errors, which may allow this
approach to reach sub m/s Doppler precision.

Another exciting possibility with this interferometer technique is to extend
RV surveys to wavelengths other than the visible, previously not covered by
echelle surveys. Since the interferometer response is simple and stable,
there is no need to calibrate the instrument response in contrast to the
echelle; only wavelength calibration is required. Hence, reference sources
with a lower line density than the iodine, which is popularly used in the
echelle, can be used. Therefore, this instrument can be easily adapted to
other wavelengths, in which more photon flux and stellar absorption lines
are available for precision Doppler RV measurements. For instance, late M, L
and T dwarfs have peak fluxes in the near-IR. Integration time can be
significantly reduced if the IR spectra can be monitored. B and A main
sequence stars and white dwarfs have very broad intrinsic absorption lines
dominated by the Balmer series. The intrinsic Doppler
error for each Balmer line is about the same as that for late type stars.
Since there are only a few dozen broad lines that can be used for early type
stars while $\sim $ 1000 lines can be used for late types, the overall
observed Doppler error is about $\sim $ 10 times higher for the early types
than late types for the same S/N data. Since early type stars are usually
much brighter in the near UV than late types at the same astronomical
distance (e.g. an A star is about 100 times brighter than a G type star in
the visible), it is possible to achieve $\sim $ a few m/s Doppler precision
by increasing S/N by a factor of $\sim $ 10.

\section {Performance of a Prototype Dispersed Fixed-delay Interferometer}

In 2000-2001, we developed a prototype dispersed fixed-delay interferometer, called
Exoplanet Tracker (ET), similar to an earlier
version built by Jian Ge and his collaborators at LLNL (Ge et al.
2002a). The instrument consists of a Michelson 
interferometer with 7 mm fixed optical path difference, a Czerny-Turner type
spectrometer with a 100 mm diameter collimator beam and a
1k$\times $1k Photonics CCD camera with 24 $\mu $m pixel size. 
The first light stellar observations  were conducted at the
Hobby-Eberly 9 m telescope (HET) in October  2001 (Ge
et al.  2002b). The spectral resolution  is $R$ = 6700. The wavelength coverage is 
140 \AA\ with the  CCD. The measured
 Doppler precision  is
approaching a photon-noise limit as demonstrated by 
a direct comparison between predicted theoretical RV errors and measured values
shown in Figure 1(left).   This figure also indicates that  the iodine calibration
contributes major errors in the measurements due to its very low
fringe visibility compared to that for the star.

\begin{figure}
\plottwo{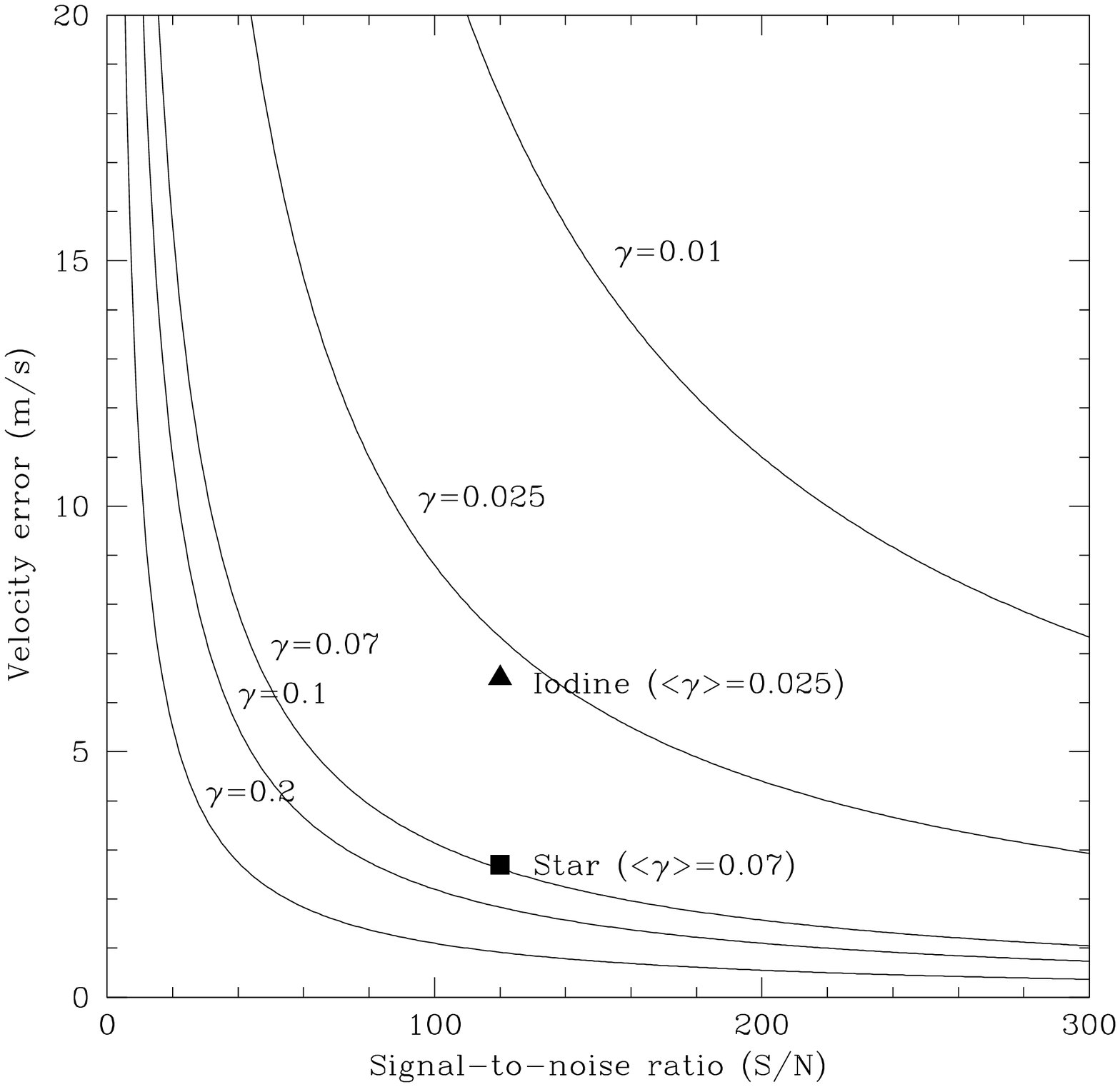}{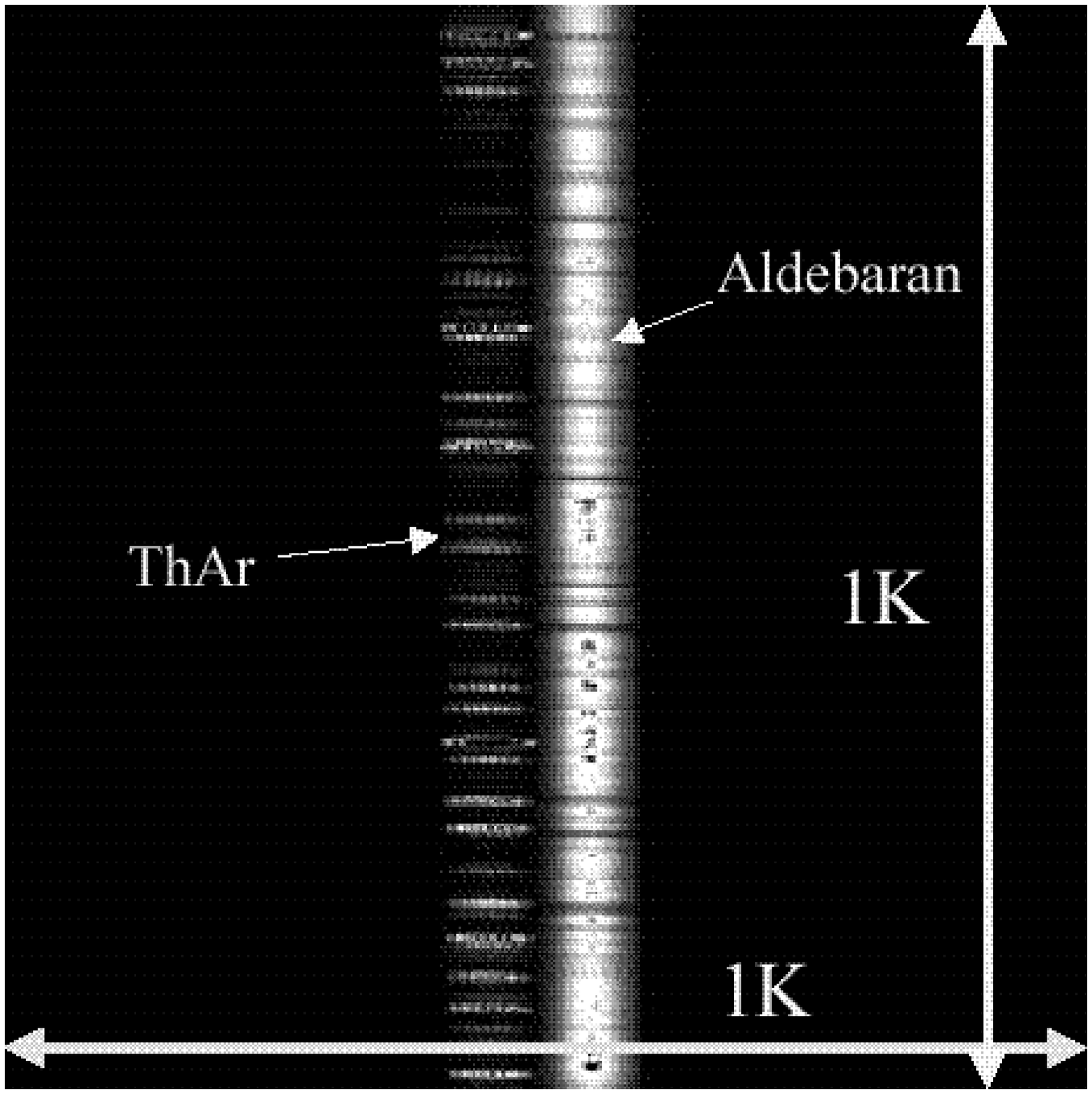}
\caption{Left: Doppler velocity error vs. S/N. The solid lines represent 
theoretical relations between RV error and S/N for different fringe visibility,
derived from Eq. (2). The filled square marks a point for the measured  error for 
a bright star, Aldebaran. 
The filled triangle represents the measured iodine error.  Right: Two 
fringing spectra, one for Aldebaran and the other for ThAr lamp,
 were simultaneously obtained with the ET and the  CCD at the 
 Palomar 5m in December 2001, demonstrating its capability 
 for multiple object observations. }
\end{figure}

ET was further tested at the Palomar 5-m telescope in Dec.
2002. Our  results demonstrate that multiple object observing is
feasible with the interferometer. Figure 1(right) shows two adjacent 
fringe spectra obtained simultaneously with  ET.   One fringe data is from Aldebaran. The other is from 
a ThAr lamp. The two fringe spectra occupy only $\sim$ 1/8 CCD 
detector area. Therefore,
we can simultaneously cover $\sim $ 15 fringe data from 15 stars 
if multiple stellar beams are available.

A modified version of  ET was used for an engineering run at the KPNO
2.1-m telescope in August 2002 before we install a
permanent one in 2003 for a long-term survey. The old $f$/10
spectrograph was replaced by an $f$/7.5 spectrograph. The KPNO 1k$\times $3k
back-illuminated CCD with 15 $\mu $m pixels was used instead of the old
1k$\times $1k CCD. The wavelength coverage has been increased to 270 {\AA}
due to the faster instrument focal ratio and larger detector array. An $f$/8
telescope beam is fed into a 200 $\mu $m fiber, which matches a 2.5 arcsec
stellar image. Due to the focal ratio degradation, the output focal ratio of
the fiber is $f$/6, which is converted to $f$/7.5 to feed the spectrograph. The
spectrograph entrance slit width was dialed to about 180 $\mu $m, which
causes about 30{\%} photon loss at the slit. The FWHM of each absorption
line is sampled by 12 pixels. Under 1.5 arcsec seeing conditions, the total
instrument throughput including the sky, telescope transmission, fiber loss,
instrument and iodine cell transmission and detector quantum efficiency is
3.4{\%}. This was achieved by feeding only one of the two interferometer
outputs into the spectrograph. This allows us to routinely observe stars as
faint as V = 7.6  during the whole run. 
Although the seeing conditions were never better than 1.5 arcsec,
 we still were able to continually monitor 6 stars (Arcturus, $\eta$
Cas, $\upsilon$ And, 51 Peg, 31 Aql, and HD 209458) over 8 nights when sky was
relatively clear. 

Part of the data have been reduced. Figure 2(left) shows a velocity
curve for $\eta$ Cas within an hour. The measured 
RV values well match the predicted ones caused by the Earth's motion. The 
RMS residual from this measurement indicates a Doppler precision of 2.9 m/s 
 in a short period. Measurements from several other
nights  indicate we have reached a Doppler sensitivity of $\sim$ 3-8 m/s for 
$\eta$ Cas. Figure 2(right) shows a RV curve from 51 Peg after Earth's motion is
subtracted, superimposed with a predicted curve from previous echelle measurements.
Our result is consistent with previous measurements, demonstrating
 that ET is capable of detecting extrasolar planets. 
Large long term RV measurement errors are mainly caused by data reduction and also 
instrument calibration.
A better version of  data reduction
software package is being developed.   Better long term RV precision
is expected.

\begin{figure}
\plottwo{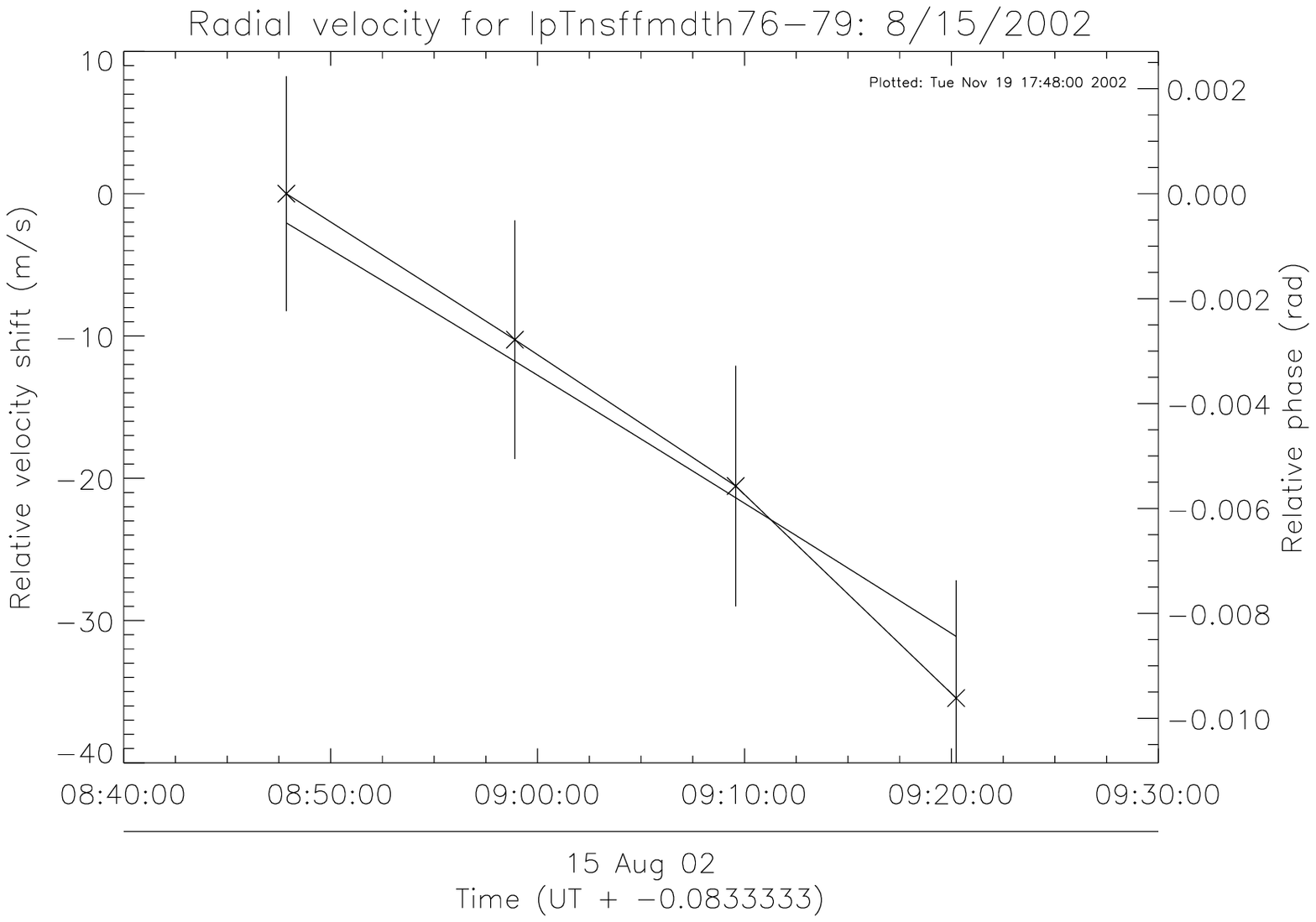}{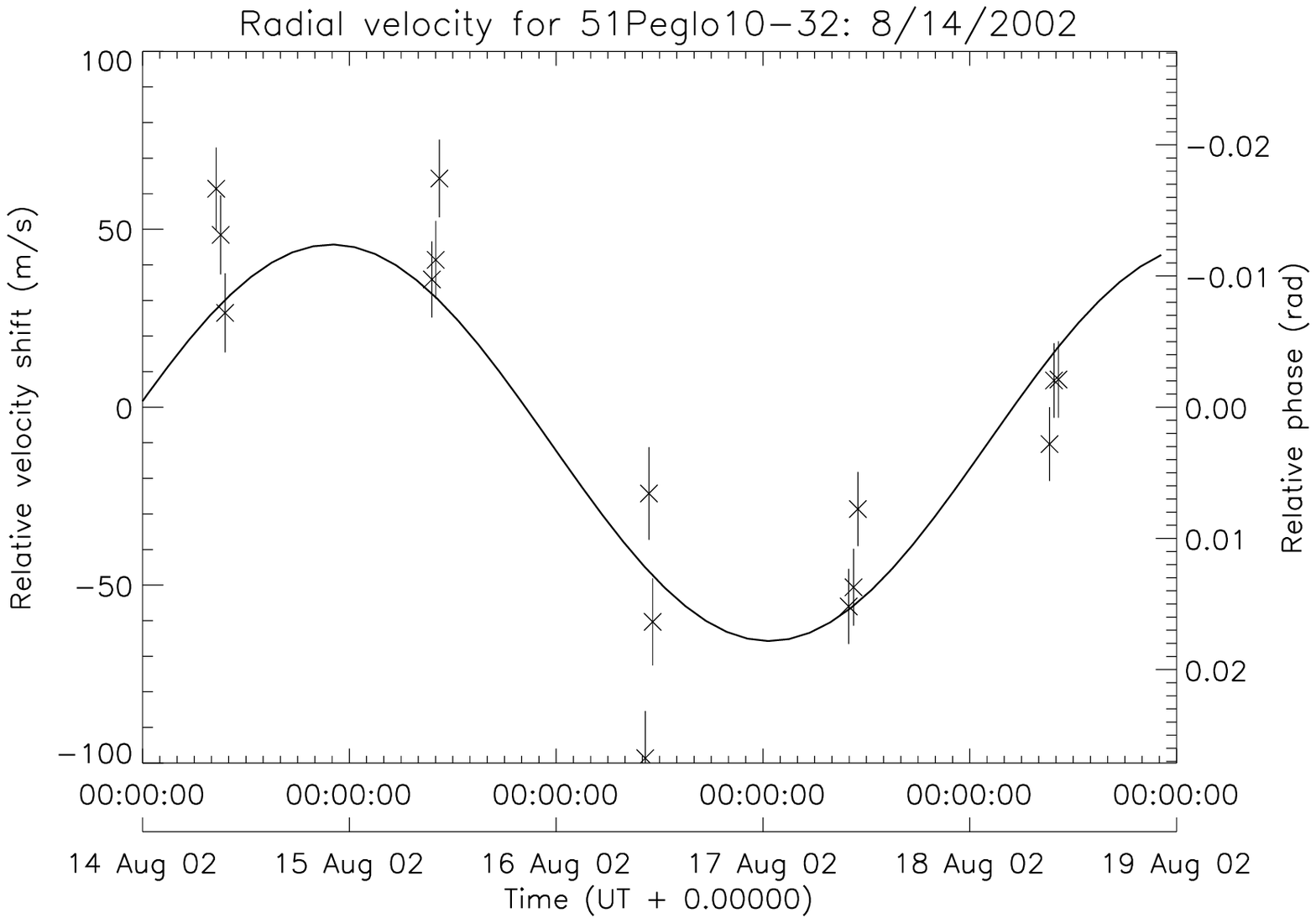}
\caption{Left: Doppler precision within an hour for Eta Cas with  ET. The RMS 
error is 2.9 m/s.   Right: Doppler velocity curve of 51 Peg  due to a planet with 
4 day period  uncovered with
ET. The solid line is the expected RV curve from previous measurements 
with echelle spectrometers. The large RV deviation in the 3rd day is partially
due to bad data calibration.   }
\end{figure}

\section {All Sky Doppler Surveys for Extra-solar Planets}

Observations with ET have demonstrated that the dispersed fixed-delay
interferometer is suitable for multiple object observing and can provide
high throughput and high Doppler precision. Multi-object and high throughput
Doppler surveys are important since they can significantly speed up the
detection of extrasolar planetary systems and also reduce the cost.
Currently, only about  4000 solar type stars are being searched by
half a dozen telescopes using the echelles. Based on current planet
detection rate of 10{\%} of solar type stars, $\sim $ 400 planets will be
discovered in the next $\sim $ 10 years. These surveys are time-consuming
and very costly. This is because current echelles can only measure a single
object per exposure, and because only relatively bright stars can be
observed since the instrument has low detection efficiency, whereas high
photon flux is required for precision measurements. Therefore, 
a multiple object RV survey can significantly  improve the current situation.

Considering there are not many bright stars in the sky for the measurements,
the multiple object survey must include relatively faint stars within the
telescope field of view (FOV). Based on our estimation from previous star
count surveys in the visible (Bahcall {\&} Soniera 1980) and a dust
extinction map (Schlegel et al. 1998), we find there are about a half
million stars from A -- M types brighter than V = 10, and about 4
millions of stars are brighter than V = 12. 
On average, about 100 stars with V magnitude brighter than 12 mag. are
within a 1 deg field-of-the-view. Therefore, in order to conduct an
efficient all sky survey for extrasolar planets with a multiple object
dispersed fixed-delay interferometer, we need to reach about V = 12  with 
a wide field telescope.

This all sky survey is possible with modern wide field telescopes such as
the Sloan 2.5-m and the WIYN 3.5m telescopes. The Sloan has a 7 degree FOV and
the WIYN has a 1 degree FOV. Since both of the telescopes are moderate
 size, high throughput becomes critical to reach high
Doppler precision for faint stars. High throughput can 
be  achieved with an optimally designed
instrument. The prototype at the KPNO 2.1-m has already demonstrated
good throughput of 3.4{\%} under 1.5 arcsec seeing. In the future, a factor
of 4 times improvement should be achievable, including 2 times in the
interferometer transmission by feeding both interferometer outputs to the
spectrograph, 1.4 times in the \'{e}tendu match through updating current
spectrograph with a faster spectrograph (e.g., $f$/2 instead of $f$/7.5), and 1.5
times in the spectrograph transmission by using a higher efficiency grating
(such as a VPH grating with $\sim $ 85{\%} efficiency instead of current
reflection grating with $\sim $ 55{\%} efficiency). A total estimated
detection efficiency of $\sim $15{\%} can be expected. 
With this detection efficiency, a Doppler precision of $\sim $ 15 m/s can be
reached for a V = 12  star within an hour integration and $\sim $ 1000
{\AA} wavelength coverage. Therefore, most of the stars brighter than V =
12 can be surveyed with high Doppler precision for planet candidates
at 2 m class wide field telescopes with multi-object RV instrument. The
candidate systems can be further studied with a single object high
throughput dispersed fixed-delay interferometer with higher precision at larger aperture
telescopes to look for additional planet members.

In the future, in order to take full advantage of the potential of the
multi-object dispersed fixed-delay interferometer for all sky RV surveys,
 we need to simultaneously feed multi-object fiber beams into
three instruments which have best sensitivity in the  near-IR,  visible  and 
near-UV, respectively. This design allows optimal match
 of the interferometer sensitivity with the peak of the spectral
flux and lines. For instance, a late M type star has about 10 times more
flux in the near-IR than in the visible and has many molecular and atomic
lines for precision Doppler measurements (Kirkpatrick et al. 1993). Because
the line width is very different from early type to very late types, the
optical delay will be changed accordingly to minimize Doppler errors. 

To
fully achieve high Doppler precision with this instrument, a large waveband
is very important since this will allow the capture of more photons from
stars for the measurements. In the prototype, the wavelength coverage is about 
200 \AA. By simply increasing wavelength coverage to $\sim$ 1000 \AA, more than 
a factor of 2 times better Doppler precision can result. 
Doppler precision can be possibly further reduced by using reference sources
with higher visibility than the iodine absorption. Our measured fringe
visibility for typical Thorium lines in our Palomar data is about 50{\%},
resulting in intrinsic Doppler error well below 1 m/s. Current RV
measurements using ThAr calibration in Dr. Mayor's group have already
achieved $\sim $ 5 m/s precision and with further improvement such as double
fiber mode scrambling and vacuum operation, they believe that they can reach
$\sim $ 1 m/s in the new HARPS high resolution echelle spectrometer (Queloz
2002 private communications; Pepe et al. 2000). We believe we may be able to
achieve similar calibration precision by using similar procedures. If this
technique works, the
observing efficiency can be tripled 
over that of our current iodine-based technique since there
will be no photon loss due to iodine absorption (typical loss by iodine
absorption $\sim $ 30-40{\%}) and also there is no need to create a
separate stellar template for each observation. This will be a big plus for
the survey.

In summary, to have a very successful all sky survey requires multi-object 
observation capability, high Doppler precision and high
throughput. A wide field telescope with at least three
multi-object dispersed fixed-delay interferometers
 optimized for near-UV, visible and near-IR
wavebands can provide the best sensitivity 
for detecting thousands of planets around stars from very early types to 
very late types in the near future.

\acknowledgements

The authors are grateful to Larry Ramsey, Don Schneider, Eric Feigelson,
Steinn Sigurdsson, Tom Soltysinski, Ron Reynolds, Fred Roesler, Didier
Queloz, Harvey Moseley, Bruce Woodgate, Roger Angel, Mike Shao, Chas
Beichman, Bill Cochran, Wes Traub, Ed Jenkins and Jim Gunn for useful
discussions. We acknowledge support from the HET staff, the Palomar
observatory staff and KPNO staff. This work is supported by NASA JPL and
Penn State Eberly College of Science. A portion of this research was carried
out at the Jet Propulsion Laboratory, California Institute of Technology,
under a contract with the National Aeronautics and Space Administration.


\begin{references}

\reference{Bahcall, J.N. and R.M. Soniera. 1980, ApJS, 44, 73}
\reference{Baranne,~A. et al. 1996, A{\&}AS, 119, 373}
\reference{Beckers, J.M., \& Brown T.M., 1978, Osser Mem Astrophys Obs Arcetri.
No. 106, 189}
\reference{Bouchy,~F., Pepe,~F., {\&} Queloz,~D. 2001, A{\&}A, 374, 733}
\reference{Butler, R.P., Marcy, G.W., Williams, E., McCarthy, C., Dosanjh,P., {\&}
Vogt, S.S. 1996, PASP, 108, 500}
\reference{Dravins, D. 1987, A{\&}A, 172, 200}
\reference{D'Odorico, S. et al. 2000, Proc. SPIE, 4005, 121}
\reference{Erskine, D.J., {\&} Ge, J., 2000, in Proc. Imaging the Universe in Three
Dimension, Edited by W. van Breugel and J. Bland-Hawthorn ASP Conference
Series, 195, 501}
\reference{Ge, J., 2002, ApJ, 571, L165}
\reference{Ge, J., Erskine, D.J., {\&} Rushford, M., 2002a, PASP, 114, 1016}
\reference{Ge, J., et al. 2002b, Proc. SPIE 4835, in press}
\reference{Gorskii, S.M., \& Lebedev, V.P., 1977, Izv Krym Astrofiz Obs 57, 228}
\reference{Kirkpatrick, D.J., et al. 1993, ApJ, 402, 643}
\reference{Kozhevatov, I.E., Kulikova, E.Kh. {\&} Cheragin, N.P. 1995, Astronomy
Letters, 21, 418}
\reference{Kozhevatov, I.E., Kulikova, E.H., {\&} Cheragin, N.P. 1996, Solar Physics,
168, 251}
\reference{Mayor, M., {\&} Queloz, D. 1995, Nature, 378, 355}
\reference{Pepe, F. et al. 2000, Proc. SPIE, 4008, 58}
\reference{Schlegel, D.J., D.P. Finkbeiner and M. Davis. 1998, ApJ, 500, 525}
\reference{Vogt,~S.S. et al. 1994, Proc. SPIE, 2198, 362}
\reference{Vogt,~S.S., Marcy,~G.W., Butler,~R.~P., Apps,~K., 2000, ApJ, 536, 902}

\end{references}
\end{document}